Position paper for 1.1 Panel on Complex Systems Engineering

# Complex Systems + Systems Engineering = Complex Systems Engineering

Russ Abbott

California State University, Los Angels
and The Aerospace Corporation

Los Angeles, Ca

Russ.Abbott@aero.org

**Abstract**
One may define a complex system as a system in which phenomena emerge as a consequence of multiscale interaction among the system's components and their environments. The field of Complex Systems is the study of such systems—usually naturally occurring, either biological or social. Systems Engineering may be understood to include the conceptualising and building of systems that consist of a large number of concurrently operating and interacting components—usually including both human and non-human elements. It has become increasingly apparent that the kinds of systems that systems engineers build have many of the same multiscale characteristics as those of naturally occurring complex systems. In other words, *systems engineering is the engineering of complex systems*. This paper and the associated panel will explore some of the connections between the fields of complex systems and systems engineering.

**What do complex systems and systems engineering have in common?**
A complex system—in the sense of the word as used to identify the field of study by that name—typically consist of elements that operate asynchronously. These elements operate according to their own schedules and are not all entrained by a single overall timing driver, i.e., a driver whose rate is transmitted throughout the system by mechanisms such as gears, drive shafts, belts, electronic timing signals, etc. Complex systems tend to exhibit what are often referred to as multiscale or emergent phenomena: phenomena at an aggregate level that (a) result from and (b) are a consequence of but (c) may be described independently of how the interactions among the individual elements bring them about. See (Abbott 2005).

Large-scale engineered systems typically consist of many concurrently operating and interacting elements—human and non-human—which as an ensemble either produce a result or provide a service or set of services. Such systems exhibit similar emergent and multiscalar properties. It has become increasingly clear that *systems engineering is the engineering of complex systems*. Although this realization is by now fairly widely shared, at least implicitly, and although complex systems ideas, tools, and techniques have been applied to systems engineering problems for some time, there has been little effort to date to bring the two fields—and the communities of people who work in these fields—together in a more formal and explicit way. The current CSER session is the first in a series of events whose goal is to introduce these communities to each other and to embark on what we expect to be an extended dialog.

Following this CSER session, sessions will be held at the IEEE System of Systems Engineering Conference (April 2006) and at the INCOSE Symposium (July 2006). A stand-alone workshop on Complex Systems Engineering is being planned for January 2007 to be followed by a full conference at which revised papers from that workshop (along with other submitted papers) will be presented to a larger audience. The follow-on conference will be sited and scheduled to coordinate with the INCOSE 2007 Symposium.

## What issues are of interest to both complex systems and systems engineering?

In forging a connection between systems engineering and complex systems, the questions to be addressed may be summarized as follows.

- *What does systems engineering need from complex systems?* What difficulties have large (and complex) engineered systems encountered that seem to derive from the fact that they are complex in the sense described above?
- *What can complex systems offer to systems engineering?* What conceptual frameworks from the field of complex systems (and the tools that result from those frameworks) can be applied to systems engineering problems?
- *What can systems engineering offer to* complex *systems?* Systems engineers have successfully built some very large and very complex systems. What intuitions have systems engineers developed about how to go about building such systems that would be of interest to the complex systems community? To what extent can systems engineering serve as a laboratory for the study of complex systems in ways that are not possible when one is studying pre-existing and naturally occurring systems?

The remainder of this paper contains an initial (alphabetical) catalogue (preliminary and tentative in many cases) of areas that we expect will be worth exploring.

### Adaptability

See *flexibility* and *evolvability*.

### Agents

Because the elements of complex systems operate asynchronously, these elements are frequently referred to as agents. Generically, an agent is an asynchronously operating element whose actions—either as a part of a system or as an actor that interacts with the system—must be taken into account when the overall operation of a complex system is being analysed.[1]

The use of an agent-based framework for analyzing or modelling a system does not in itself imply a particular level of agent autonomy or intelligence. Agents may be minimally autonomous and may simply follow orders—although being agents whose schedule is not entrained by whatever is generating the orders, they follow their orders according to their own scheduling mechanism. Alternatively, so-called *intelligent agents* may be understood as having significant autonomy and intelligence and as capable of being given a goal and then being set loose to achieve that goal using some combination of allocated and found resources.

In both complex systems and systems engineering agent-based modelling allows one to explore how behaviors defined at the level of system elements (the agents) give rise to phenomena at the system level. (See *emergence*.)

An agent-based perspective is important in a second way for systems engineering: it defines what is potentially a common conceptual framework for system architectures. Just as the language of object-oriented programming has become the *de facto* vocabulary for large scale soft-ware systems, the language of agent-based architectures are likely to become the *de facto* vocabulary for engineered complex systems.

---

[1] See, for example, Tesfatsion's web site, available as of January 31, 2006 at http://www.econ.iastate.edu/tesfatsi/ace.htm, for a fairly comprehensive list of agent-based modeling resources.

See also *distributed control*, *market mechanisms*, and *stigmergy*.

**Architectures**

For the most part, members of the complex systems community do not think in terms of system architectures in the same way that systems engineers do.[2] This is not to say that architectures are not important in understanding naturally occurring complex systems. Scientists who study biological, ecological, and social systems work hard to understand their underpinnings and internal organizations. But systems engineering-style architectures sense seem not to be central.

Are there systems engineering architectural concepts that can be helpful in explicating the operation of naturally occurring complex systems? If so, this would be an area in which intuitions developed by systems engineers can enrich the field of complex systems.

On the other hand, if systems engineering-style architectures turn out not to be useful for understanding naturally occurring complex systems, one must wonder why these frameworks have value for systems engineering and not for complex systems. One possible reason might be that systems engineers are responsible for building systems whereas complex systems scientists study systems that already exist. Does development responsibility require a level of architectural support that isn't needed when simply studying a system?

**Bricolage**

*Bricolage* is the use of whatever is at hand to satisfy an immediate need—as when a rock is used as a hammer. Reflecting a bricolage-style design, many naturally occurring systems include components that support a wide range of uses. Hands can be used to type, to assess temperature and roughness, to click a mouse, to grasp, to sew, to strum, to point, to caress, to scratch, and to strike. Faces can express emotions; they can also ingest air, food, and light waves. In naturally occurring systems, once a second use for a system component has established its value, evolution tends to enhance its usability for that purpose. To what extent is bricolage acceptable/desirable as a systems engineering design approach?

See also *system of systems*.

**CONOPS (Concept of Operations)**

A system's CONOPS is the plan for how it will be used once deployed. What is lacking from many CONOPS considerations is a realization that the user is outside the system and is embedded in a larger system, which may put pressure on him/her to use the system in unexpected ways.

See *systems of systems*.

**Distributed control**

Because complex systems consists of agents, the functionality produced by a complex system is a by-product of the interaction among the agents in the system. Desirable as it may be to be able to point to a single place in a system as the ultimate determiner of how the system will behave, the actual behavior of a system depends on all its components, which, being agents, are not under the entrained control of a single system authority. Systems with distributed control are common in both naturally occurring and engineered complex systems.

---

[2] See, for example, the various architecture frameworks (such as DODAF and Zachman) described on the Systems and Software Consortium's *Architecture* web page, available as of January 31, 2006: http://www.software.org/pub/architecture/fwhome.asp. Consider also the various specification languages and meta-languages (such as UML) produced by the Object Management Group. See their web page, available as of January 31, 2006: http://www.uml.org/.

It is possible, of course, to design a system with agents that have absolutely minimal autonomy—or to design a system in which there are no agents at all and in which the entire system is understood as a single immensely complicated device. Systems are not designed in this manner.

1. In a large system there are so many micro-level decisions to be made that to make them all at a central decision making authority would turn that central authority into an enormous bottleneck.
2. Were there a central decision making authority, it simply would not have the expertise and detailed information to make all the required decisions intelligently.

As experience with command-driven economies illustrate, such rigidly hierarchical and micro-managed systems are both inefficient and ineffective.

A system that is rigidly micro-managed has an additional significant weakness: its decision making process is vulnerable to manipulation. Two sorts of vulnerabilities occur, depending on the style with which decisions are made.

1. If decisions tend to be based on clearly specified rules and regulations, entities that need to have decisions made learn how to manipulate requests so that they will get the results they want. This is frequently called *gaming the system*: submit a request formulated in a particular way because one knows how the rules will require the system to respond.
2. If decisions are made by human beings with broad discretionary powers, manipulate the person—rather than the rules—either through a bribe or through some less overt form of pressure, entreaty, or temptation. This reflects an alternative gambit: if one doesn't succeed in accomplishing one's end within a particular system, retreat to the larger system within which the subject system is embedded. This is discussed further in (Abbott 2006).

The fact that a centralized decision making authority is vulnerable to manipulation confirms the inevitable agent-based nature of any truly complex system. As we have seen, even in a system in which one attempts to centralize decision making, many of the real decisions are in fact made elsewhere. The primary difference is that an additional burden of manipulative overhead is imposed on the agents—and hence on the system as a whole.

See also *agents*, and *systems of systems*.

**Energy**

Both naturally occurring and engineered complex systems must accommodate themselves to the fact that neither energy nor mass is emergent. Although they can be interchanged, neither can be created from interactions among more primitive elements. In most naturally occurring and engineered complex systems mass tends to be fixed.[3] Thus energy is critical.

Both naturally occurring and engineered complex systems persist and function only in so far as they are powered by external sources of energy. (Hence the frequent reference to "far from equilibrium" systems.) Naturally occurring complex systems are typically powered by sunlight or its stored equivalents. Social complex systems are typically powered by energy proxies such as money. Engineered complex systems also require sources of energy. We hypothesize that analyses of energy flows will reveal fundamental similarities among naturally occurring and engineered complex systems. "Follow the money" really means "follow the energy."

---

[3] Any release of energy involves the conversion of some mass to energy. But other than in the release of nuclear energy, the amount of mass involved is negligible.

**Emergence**
    See *requirements*.

**Engineering**
    See *science*.

**Environments**
    See *systems of systems*

**Evolutionary computation**

Evolutionary computational techniques (genetic algorithms and genetic programming) have been applied to engineering problems since they were first developed. Recent GECCO conferences feature what are called *human competitive* (Koza 2003) designs produced through evolutionary means.[4] A somewhat more recent development involves work in a relatively new subfield of complex systems known as developmental systems (Hornby 2007).

We are awaiting the time when computational power and sophistication will be sufficient to enable agents in agent-based models (or even in agent-oriented system implementations) to model the worlds in which they are functioning and to perform their own on-the-fly genetic programming computations in order to determine their most effective courses of action.

**Evolvability**

Evolution is the first principle of biology. It is desirable for complex systems to have the structural capacity to evolve. Yet a fundamental difference between systems that are engineered and systems that result from natural processes is that engineered systems are often optimized for one or more particular properties with little thought given to evolvability. When the environment in which an engineered system changes or when the uses to which one wants to put an engineered system changes, it is often quite difficult to change the system to accommodate those new needs. Natural systems tend to be more adaptable. Can we understand how this difference comes about and what, if anything we can do about it?

It seems (a) intuitive, (b) well established theoretically (Adami 2000) (Heylighen 1996), and (c) obvious just by looking around that evolution leads to an increase in complexity. Yet the increasingly complex systems that evolve are not re-designed from scratch with every evolutionary step. Each step is an evolutionary neighbor of something a bit less complex. The only way such an evolutionary process can occur is if the evolutionary sequence provides both (a) the specific features needed at each step and (b) a framework (or architecture) that can support the evolutionary process itself. In other words, the only architectures that survive an evolutionary process are those that support and facilitate evolutionary change. Thus the evolutionary process produces not only fitness for changing environments but evolvability itself.

Successive versions of most engineered systems are not forced to run this sort of evolutionary gauntlet. New versions of many engineered systems are often built from scratch; they are not evolved as extensions of previous versions. And those that are extensions of previous versions are often quite clumsy. Of course many evolved systems are also quite messy. But in the long run, evolution prunes away structures that impede evolution leaving evolvable core frameworks.

---

[4] The "Human-Competitive Results" page of the genetic-programming.com web site (available as of January 31, 2006: http://www.genetic-programming.com/humancompetitive.html) lists 36 human-competitive results as of December 31, 2003. See also http://www.genetic-programming.org/hc2005/cfe2005.html (available as of January 31, 2006) for the 2005 GECCO "Humie" awards.

We fail to build evolvable systems because we often don't try. Evolvability is not required of most engineered systems. One reason for this is that designing for evolvability is easier said than done. A second is that we don't have a satisfactory way either of specifying or measuring evolvability. Given the current state of our knowledge about evolvability, the most effective way of requiring evolvability is economic: require developers to absorb some of the cost of post-delivery enhancements. It's difficult to see, however, how one would structure a contract to that effect that was attractive to developers and fair to both developers and customers.

On the other hand, we as system builders do learn from previously built systems. Although we don't build systems that themselves are evolvable, we do evolve our own knowledge of how to build better and better systems. The analogy is that an individual's DNA is comparable to the initial conceptualization of a system; a fully grown and functioning individual is comparable to a functioning system; and the developmental process, in biology known as ontogeny, which we now know is controlled by gene switches (Carroll, 2005), is comparable to our retained and evolved knowledge of how to build systems. Just as we now know that as much of biological evolution is in the gene switches (and not in the genes), much of the evolution (retained knowledge) of systems engineering is in the theory, tools, and practices of development.

**Failures**
See *power laws*.

**Far from equilibrium**
See *energy*.

**Genetic algorithms and genetic programming**
See *evolutionary computation*.

**Market mechanisms**
Market-based mechanisms allow agents to interact within a defined framework to exchange goods, services, and resource proxies (such as money). An important advantage of a market mechanism is that the decision making process is completely transparent and (mainly) immune to manipulation. Market mechanisms frequently occur in both social and engineered complex systems and are important to both complex systems and systems engineering.

A market is the visible part of an iceberg whose hidden part consists the rest of the economic system within which the market functions. A market optimizes the use of resources by setting resource rates of exchange that reflects the relative values of the resources in the larger context. Markets work only because the resources traded have value to the market participants outside a market. Otherwise, it's not clear what benefit the participants would gain from participating in the market. If a market were an isolated (and sterile) mechanism for performing exchanges, each participant would be exactly as well off after a trade at the market price as he or she was before.

Markets work only when each participant is better off after a transaction than before. *Better off* means that the resources they have after the transaction are worth more *to them* than the resources they had before the transaction. Why would a resource be worth more to one market participant than to another? The answer is that it is the differences in the situations of the market participants outside the market that makes markets work. A simple example is the sale of an electronic component. The seller manufactures the component for less than he receives in the market. That's why he participates. The buyer uses the component in a product that he sells for more than the sum of the costs of the components and the cost of assembling them. That's why he participates. It is only because each participant is situated differently in the larger system that they value the component differently and are each better off after trading in the market.

When a collection of agents, each of which is individually and differently situated in a larger context, can each place a value on various resources from the perspective of their own situations, then a market provides a mechanism to reconcile those various values and to optimize the use of those resources. Markets don't set a price. They allow exchange rates to adjust as perceived external values change. Other extra-market properties (e.g., inventories, financial resources, market impact) may also affect buy/sell decisions and thereby the prices at which resources are traded.

See also *agents*.

**Modeling**

See *agents*.

**Non-algorithmic programming**

See *requirements*.

**Non-linearity**

Non-linearity does not just mean more sophisticated mathematics.
See *phase transitions*

**Phase transitions**

The operation of many systems varies relatively smoothly over a wide range of the system's parameters. A phase transition is a point at which a small change in a system parameter causes a very large change in the performance, nature, or operation of the system. Many engineered and naturally occurring complex systems exhibit phase transitions—often unanticipated.

**Power laws**

Failures in the electrical power grid follow a power law when intensity is plotted against frequency. (Talukdar 2003) Similar power laws are observed in a wide range of phenomena—intensity vs. frequency of earthquakes (Christensen 2002) being probably the best known.

It is hypothesized that failures in engineered systems—when the cost or significance of the failure is plotted against the frequency—will also be found to follow a power law. It is also hypothesized that this is an optimisation effect for which (Doyne 2005) coined the terms *robust yet fragile* and *highly optimized tolerance*. As we understand this phenomenon better, the insights we gain may be applied when assessing risk, fault tolerance, and other failure-related issues.

**Requirements**

(Abbott 2005) defines a property or phenomenon as emergent if it can be specified independently of its implementation. Specifying system properties and phenomena in this way is fundamental to systems engineering. Emergence seems somewhat more mysterious in naturally occurring complex systems because it occurs without apparent intention.

We find it mundane to engineer a satellite in geosynchronous orbit because we know how to do it. Yet the specification of geosynchronicity—that a satellite remain motionless with respect to earth as a reference frame—is specified independently of its implementation. One might naively imagine achieving such an effect by tying a long cable to the satellite, as if it were a balloon. Of course that's not how it works. Geosynchronicity is achieved by putting the satellite in an orbit whose period is the same as the period of the earth's rotation. There is nothing about this (or any particular orbit) that holds a satellite motionless with respect to earth as a reference frame. Geosynchronicity emergences as a consequence of the fact that the two periods are the same. One might refer to this as planned emergence or non-algorithmic programming.

When ants forage or birds flock as a result of relatively simple rules we may (initially) find this amazing—because the emergent phenomena seem to appear as if by magic from a system of components which are following rules that on their face seem to have nothing to do with the end result. But once we understand (Spector 2003) how these phenomena result from the underlying rules, this sort of emergence is no more (and perhaps no less) mysterious than geosynchronicity.

Two other everyday examples of naturally occurring emergence are the 24-hour day-night cycle and the yearly seasonal cycle. The first is an emergent consequence of the earth's rotation with respect to the sun as a source of light. The second is an emergent consequence of the tilt of the earth's axis and its revolution about the sun as a source of heat. In systems engineering one often starts with requirements, which are emergent properties of a hypothetical system. One then attempts to reverse engineer such a hypothetical system to determine how such emergence might actually be accomplished. If one didn't know about the geometric relationships between the earth and the sun, how might one engineer a system that exhibited day-night and seasonal cycles?

**Reuse**

See *architectures* and *evolvability*.

**Risk**

See *failures*.

**Science**

Engineering is the application of scientific and mathematical principles to practical ends. Yet there is no *science* to which systems *engineering* looks for its principles. Its body of knowledge consists mainly of best practices. Complex systems is the science that studies, among other things, the sorts of systems that systems engineers build. To the extent that complex systems establishes itself as a science, it can serve as one of the sources of scientific principles that systems engineering, as an engineering discipline, can apply.

**Service Oriented Architecture (SOA)**

An architecture that organizes a system as a framework within which capabilities are made available as services. This is much the same model as the economy in which products and services are made available to those who might (buy and) use them. See *system of systems*.

**Stigmergy**

(Bonabeau 1999) borrowed the term *stigmergy* from entomology to refer to the use of the environment for communication among agents in an agent-based systems. Probably the most familiar example is ant foraging in which ants leave pheromone trails (in the environment) for other ants to follow. Most engineered systems tend not to use this kind of mechanism and to favor instead communication mechanisms over which the system has more direct control.

See *systems of systems*.

**Systems of systems**

A system that never changes—even if its components are themselves systems—is just another system. In (Abbott, 2006) we characterize the nature of systems of systems in terms of the following properties. A system of systems is (a) open at the top (to new applications), (b) open at the bottom (to new implementation of primitives), and (c) continually but slowly changing.

An additional and essential aspect of a system of systems perspective is the recognition that every system, other than nature itself, is embedded in an environment—which is a larger system.

A fundamental insight—and one that ties work in systems of systems to work in complex systems—is that whenever something exists in an environment, other elements of that environment may use it in ways that were not originally anticipated.

A fundamental example is the predator-prey relationship. It may sound trivial to say so, but predators are able to exploit their prey only because the prey exist in the predator's environment. But the prey typically[5] don't place themselves in the environment for the benefit of the predators.

Exploitation of an existing element in an environment isn't always a one-way street. Plants and bees are the standard example of mutual (and mutually beneficial) exploitation.[6, 7] Each makes use of the existence of the other for its own purposes. The relationship between them has become so intimate that each now depends for its survival on the use the other makes of it.

The fundamental principle is that once a system (or any mechanism) exists in the world, it may be used for purposes that are different from those intended or anticipated when the system or mechanism was created and deployed. This is a source of both great creativity and unintended consequences. Anti-bacterial cleaners have led to the evolution of bacteria that are resistant to the anti-bacterial agents in the cleaners. Tax laws have led to an accounting specialty that finds ways to avoid the effects of those laws. In both cases, the creation of a mechanism in the world that is noxious to a population has led to the development of counter-mechanisms that allow elements of the population to escape the most noxious aspects of those mechanisms.[8]

To build a successful system of systems, one must adopt an environmental—one might call it an outward facing—perspective. Such a perspective leads to the following questions which must be addressed about any system—and especially about systems of systems.

- On what environmental presumptions/capabilities does the system depend? In what ways does the system's functioning depend on environment-based (i.e., stigmergic) interactions? Can the system make more use of stigmergic mechanisms?
- How does the system change the environment within which it exists simply by existing? How do/might non-system elements interact with/exploit/foil it once it exists? What symbiotic relationships exist among the system and other elements of the environment? Are these desirable/inevitable?
- How does the system serve as an environment for its components?
- If the system is to exist within an environment that is undergoing rapid evolution, how can one build it so that it won't be obsolete by the time it's finished? How can it be designed so that it can evolve as the environments within which it exists evolves around it?

Human civil society is a system of systems. We add new systems to it all the time, and it continues to function—more or less. What is it about civil society that enables it to integrate more and more systems into itself without apparent limit—at least so far?

See also *agents*, *distributed control*, *market mechanisms*, and *stigmergy*.

## References
Abbott, R, "Emergence Explained," Submitted for publication, October 2005. Available as of January 31, 2006: http://cs.calstatela.edu/~wiki/images/9/90/Emergence_Explained.pdf.

---

[5] An interesting example of prey actually inserting themselves into an environment for the benefit of predators is the strategy adopted by a team from the University of Southampton in a Prisoner's Dilemma tournament (Grossman, 2006). Some Southampton entries sacrificed themselves for the benefit of other Southampton entries, which won the tournament—beating Tit-for-Tat for the first time.

[6] Here *exploitation* is intended to be understood as "to make use of" and not necessarily to disadvantage another.

[7] An environment that includes an interacting "plants system" and "bees system" is a system of systems.

[8] The "Museum of unintended consequences" is an informal web collection of such unintended consequences. It is available as of January 31, 2006: http://cs.calstatela.edu/~wiki/index.php/Courses/CS_461/Museum_of_unintended_consequences

## Biography

Russ Abbott is a Professor of Computer Science at California State University, Los Angeles and a member of the staff at The Aerospace Corporation.